\documentclass[runningheads]{llncs}
\usepackage[T1]{fontenc}
\usepackage[utf8]{inputenc}

\usepackage{graphicx}

\usepackage{amsmath}	
\usepackage{amssymb}
\usepackage{fncylab} 
\usepackage{stmaryrd}
\usepackage{color}
\usepackage{url}
\usepackage{tikz}

\newtheorem{thm}{Theorem}

\newtheorem{prp}[thm]{Proposition}
\newtheorem{lem}[thm]{Lemma}
\newtheorem{cor}[thm]{Corollary}
\newtheorem{ex}[thm]{Example}
\newtheorem{df}[thm]{Definition}
\newtheorem{clm}{Claim}
\newcommand{\bthm}{\begin{thm}}
	\newcommand{\bprp}{\begin{prp}}
		\newcommand{\bcor}{\begin{cor}}
			\newcommand{\blem}{\begin{lem}}
				\newcommand{\bex}{\begin{ex}\normalfont}
					\newcommand{\bdf}{\begin{df}\normalfont}
					\newcommand{\ethm}{\end{thm}}
					\newcommand{\eprp}{\end{prp}}
				\newcommand{\ecor}{\end{cor}}
			\newcommand{\elem}{\end{lem}}
		\newcommand{\eex}{\qed\end{ex}\normalfont}
	\newcommand{\edf}{\end{df}\normalfont}
\newcommand{\bp}{\begin{proof}}
	\newcommand{\ep}{\qed\end{proof}}

\newcommand{\ra}{\rightarrow}

\labelformat{enumi}{(\roman{enumi})}

\newcommand{\BA}{BA}
\newcommand{\DBA}{DBA}
\newcommand{\TA}{TA}
\newcommand{\DTA}{DTA}
\newcommand{\MDBA}{MDBA}

\newcommand{\FUDT}{\mathcal{FU\text{-}DT}}

\definecolor{ao}{rgb}{0.0, 0.7, 0.0}

\begin{document}
\title{Deciding Top-Down Determinism of\\ Regular Tree Languages}
\author{Peter Leupold\inst{1}
\and
Sebastian Maneth\inst{1} 
	\authorrunning{P. Leupold and S. Maneth}
	%
\institute{Faculty of Informatics, Universität Bremen, Germany\\
		\email{leupold/maneth@uni-bremen.de}
}
}	%
\maketitle             
\begin{abstract}
It is well known that for a regular tree language it is decidable
whether or not it can be recognized by a deterministic top-down
tree automaton (\DTA). However, the computational complexity of this problem
has not been studied. We show that for a given deterministic bottom-up
tree automaton it can be decided in quadratic time whether or not its
language can be recognized by a \DTA. Since there are finite tree languages
that cannot be recognized by \DTA{}s, we also consider finite unions
of \DTA{}s and show that also here, definability within deterministic bottom-up
tree automata is decidable in quadratic time.
\end{abstract}

\section{Introduction}
Unlike for strings, where left-to-right and right-to-left deterministic automata recognize
the same class of languages, this is not the case for deterministic tree automata:
deterministic top-down tree automata (\DTA{}) only recognize a strict subset of the
regular tree languages. 
The most notorious example of a tree language that cannot be recognized by \DTA{} is
the language $\{f(a,b),f(b,a)\}$. Nevertheless, \DTA{} bear some advantages over
their bottom-up counterpart: they can be implemented more efficiently, because a tree
is typically represented top-down and identified by its root node (also, a \DTA{} may
reject a given tree earlier than a bottom-up tree automaton). 

Several properties have been defined that characterize \DTA{} within the regular tree languages.
Viragh~\cite{bViragh81} proves that the regular, ``path-closed'' tree languages are 
exactly the ones that are recognized by \DTA. He proves this via the construction of what he 
calls the powerset automaton for the path-closure of a regular language.
Gécseg and Steinby use a very similar method in their textbook~\cite{bTreeAutomataBook}.
Another approach is Nivat and Podelksi's homogeneous closure~\cite{bNivatPodDET}. 
Also here the tree automaton constructed for the closure has as state set the
powerset of the original state set.
In neither case an exact running time has been investigated.

In our approach, starting from a given deterministic bottom-up tree automaton, we first
construct an equivalent minimal automaton. This takes quadratic time, following well known
methods. We then lift the ``subtree exchange property'' of Nivat and Podelski~\cite{bNivatPodDET}
to such an automaton; essentially it means, that if certain transitions are present,
e.g., $f(q_1,q_2)\to q$ and $f(q_2,q_1)\to q$, then also other transitions \emph{must}
be present (here, also $f(q_i,q_i)\to q$ for $i=1,2$).
This property characterizes the \DTA{} languages and can be decided in linear time.
Finally, if the decision procedure is affirmative, we show how to construct
an equivalent deterministic top-down tree automaton. 
The construction replaces so called ``conflux groups'' (e.g., the four transitions from above),
one at a time by introducing new states. Care has to be taken, because the removal of one
conflux group may introduce new copies of other conflux groups. However, after all original 
conflux groups are eliminated, the removal of newly introduced conflux groups does not cause
new conflux groups to be introduced. 
We then generalize our results to \emph{finite unions of deterministic top-down 
tree languages}. We show that they are characterized by minimal bottom-up tree automata
where a finite number of ``violations'' to the above exchange property are present. 
This finiteness test can be achieved in linear time.

For unranked trees several classes of deterministic top-down tree languages have
been considered. For all of them, the decision whether a given unranked regular
tree language 
belongs to one of these classes takes exponential time~\cite{bSimplifyXMLSchema,bComplexityXMLSchema,bPhDMartens}. 
This is in sharp contrast to our results.
The reason is that the unranked automata use regular expressions in their rules,
and that inclusion needs to be tested for these expressions.

\section{Preliminaries}

{\bfseries Trees.}\quad
   For a {\em ranked alphabet} $\Sigma$ 
   we denote by $\Sigma_k$ the set of all symbols which have rank $k$.
   Let $X = \{x_1,\dots \}$ be a set of constants called {\em variables};
   for an integer $n$ we denote by $X_n$ the set $\{x_1,\dots, x_n\}$ of $n$ variables.  
   The set $T (\Sigma, X )$ of {\em trees over the ranked alphabet} $\Sigma$ and the set $X$ of variables is the smallest set defined by:
   \begin{itemize}
   	\item $\Sigma_0 \subseteq T (\Sigma, X )$,
   	\item $X \subseteq T (\Sigma, X )$, and
   	\item if $k \geq 1$, $f \in \Sigma_k$ and $t_1 ,\dots , t_k \in T (\Sigma, X )$, then $f (t_1 ,\dots , t_k ) \in T (\Sigma, X )$.
   \end{itemize}
We denote by $T(\Sigma)$ the set of trees in $T (\Sigma, X )$ which do not contain variables.

For a tree $t=f(t_1,\dots,t_k)\in T(\Sigma,X)$ we define its {\em set of nodes} as
$$N(t) :=\{ \epsilon\} \cup \{iu \mid i\in\{1,\dots,k\}, u\in N(t_i)\}.$$
Here $\epsilon$ denotes the root node.
Let $t\in T(\Sigma,X_n)$ and $t_1,\dots,t_n \in T(\Sigma,X)$.
Then $t[x_1\leftarrow t_1, \dots, x_n\leftarrow t_n]$ denotes the tree obtained from $t$ by replacing each occurrence of $x_i$ by $t_i$.

\noindent    
{\bfseries Tree Automata and Transducers.}\quad
A {\em (bottom-up) tree automaton} (\BA) is a tuple $A = (Q, \Sigma, Q_f , \delta)$ where
$Q$ is a finite set of states, $Q_f \subseteq Q$ is a set of final states, and $\delta$ is a set of transition rules of the following form:
$$f (q_1 , \dots , q_k ) \ra q,$$
where $k \geq 0$, $f \in \Sigma_k$, and $q$, $q_1$ , $\dots$ , $q_k \in Q$. 

A tree automaton is {\em deterministic} (\DBA) if there are no two rules with the same left-hand side.
By $A(t)$ we denote the unique state that is reached in a deterministic bottom-up tree 
automaton by processing the tree $t$.
For a bottom-up tree automaton $A$, by $A_q$ we denote the same automaton just with $q$ as the single final state, that is $Q_f=\{ q \}$.

A {\em top-down tree automaton }  (\TA)
 is a tuple $A = (Q, \Sigma, I, \delta)$ where $Q$ is a set of states, $I \subseteq Q$ is a set of initial states, and $\delta$ is a set of transition rules of the following form:
$$q(f) \ra f (q_1 , \dots , q_k ),$$
where $k \geq 0$, $f \in \Sigma_k$, and $q$, $q_1$ , $\dots$ , $q_k \in Q$. 
A top-down tree automaton $(Q, \Sigma, I, \delta)$ is {\em deterministic } (\DTA{}) if there is
one initial state and there are no two rules with the same left-hand side.

A run of a BA on a tree $t$ is a mapping $\beta: N(t) \rightarrow Q$ which fulfills the following properties: 
for all nodes $u\in N(t)$, if $u$ of rank $k$ has label $f$, $\beta(u) = q$ and for all $i\in\{1,\dots,k\}$ we have $\beta(ui) = q_i$, then
$f(q_1,\dots,q_k) \rightarrow q$ is a transition in $\delta$.
We denote the transition that corresponds to $\beta(u)$ by $\tau(\beta(u))$.
Sometimes we will view $\beta$ as a tree and refer to nodes $\beta(u)$; here we mean a relabeling of $t$ where every node $u$ is labeled by $\beta(u)$.

For a bottom-up tree automaton $A$ the run $\beta$ recognizes the tree $t$ if $\beta(\epsilon) \in Q_f$.
A tree is recognized by $A$ if there exists an accepting run for it.
The language recognized by the automaton denoted by $L(A)$ is the set of all trees which are recognized.
A tree language is {\em regular}, if it is recognized by some bottom-up tree automaton.

For an \BA{} $A$ its {\emph{corresponding}} \TA{} $c(A)$ is obtained by reading $A$'s transitions from right to left and taking $A$'s final states as initial states. 
In the same way for a \TA{} its \emph{corresponding} \BA{} is defined. 
The language of a \TA{} $B$ is defined as $L(c(B))$.

\noindent
{\bfseries Syntactic Congruence.}\quad \label{sSynCon}
	A tree $C \in T (\Sigma, X_1 )$ is called a {\em context}, if it contains exactly one occurrence of the variable $x_1$.
	Because there is only one fixed variable, we write $C[t]$ instead of $C[x_1\leftarrow t]$.
	We denote by $\mathcal{C}(\Sigma)$ the set of all contexts. 
	For a given tree language $L$ we define the syntactic congruence $\equiv_L$ on $T(\Sigma)$ by: $s\equiv_L t$ if for all contexts $C\in \mathcal{C}(\Sigma)$ we have $C[s]\in L$ iff $C[ t]\in L$.
	
	In the case of string languages the Myhill-Nerode-Theorem states that a language is regular if and only if its syntactic congruence is of finite index~\cite{bMyhill,bNerode}. 
	An analogous result exists for tree languages and was long regarded as folklore; Kozen explains its history and provides a rigorous proof~\cite{bKozenMNTrees}.
	
	A concept closely related to the syntactic congruence is the {\em minimal deterministic bottom-up automaton} (\MDBA). It is defined as follows: 
	Let $Q$ be the finite set of equivalence classes of $\equiv_L$ for a language $L$ minus the unique equivalence class $C_\bot$ of all trees $t$ for which there does not exist any context $C$ such that $C[ t] \in L$. 
	We denote by $[t]$ the equivalence class of a tree $t$ and define the transition
	function $\delta$ by:
	$\delta (f ([t_1 ], \dots , [t_k ]) = [f (t_1 , \dots , t_k )]$
	for all $t_1,\dots,t_k\in T(\Sigma)\setminus C_\bot$ and $[f (t_1 , \dots , t_k )] \not = C_\bot$.
	With $Q_{f} = \{[u] \mid u \in L\}$  the \DBA{} $M_L := (Q , \Sigma, Q_{f} , \delta)$
	recognizes the tree language $L$.
	So the states of the \MDBA{}  for a language correspond to the equivalence classes of the syntactic congruence~\cite{bTata2007}.

\bprp\label{MDFA}
Let $L\subseteq T(\Sigma)$ and $M=(Q,\Sigma,Q_f,\delta)$ be the corresponding MDBA.
Then the following properties hold.
\begin{enumerate}
	\item\label{MDFA1} For all $q\in Q$ the language $L(M_q)$ is not empty,
	\item\label{MDFA2} every transition in $\delta$ is useful, i.e., it is used in some accepting run,
	\item\label{MDFA3} for all $t\in T(\Sigma)$ we have $|\{q\in Q \mid t\in L(M_q)  \}| \leq 1$.
\end{enumerate}
\eprp
\ref{MDFA1} holds because the syntactic congruence does not have empty classes.
For every tree, which is not in the class $C_\bot$ there exists a context $C$ such that $C[x_1\leftarrow t] \in L$ by the definition of the equivalence classes, which proves~\ref{MDFA2}.
$M_L$'s determinism has~\ref{MDFA3} as a direct consequence.

\noindent
{\bfseries Subtree Exchange Property.}\quad \label{sExch}
The class of all languages that are recognized by \DTA{}s is defined via these automata. 
However, there are several other characterizations by different means.
An early one that later became known as the {\em path-closed} languages was provided by Viragh~\cite{bViragh81}.
The path language $\pi(t)$ of a tree $t$, is defined inductively
by:
\begin{itemize}
	\item if $t \in \Sigma_0$ , then $\pi(t) = t$
	\item if $t = f (t_1 ,\dots , t_k )$, then $\pi(t) = \bigcup^{i=k}_{i=1} \{fiw\mid w\in \pi(t_i)\} $
\end{itemize}
For a tree language $L$ the path language of $L$ is defined as $\pi(L) = \bigcup_{t\in L} \pi(t)$, the
path closure of $L$ is defined as $pc(L) = \{ t\mid \pi(t) \subseteq \pi(L)\}$.
A tree language is path-closed if $pc(L) = L$.
Viragh proved that the regular, path-closed tree languages are exactly the ones that are recognized by deterministic top-down automata. 
Nivat and Podelski argued that in these languages it must be possible to exchange certain subtrees~\cite{bNivatPodDET}.
We will extensively use this so-called exchange property in a formulation by Martens et al.\cite{bDETPastPresFut}.

\bdf\label{bExchange}
A regular tree language $L$ fulfills the {\em exchange property} if, for every $t \in L$ and every node
$u \in N(t)$,
if $t[u \leftarrow f(t_1 , \dots, t_k )] \in L$ and also $t[u \leftarrow f(s_1 , \dots , s_k )] \in L$, then
$t[u \leftarrow f(t_1 , \dots ,t_{i-1}, s_i, t_{i+1} , \dots , t_k )] \in L$ for each $i = 1, \dots , k$.
\edf
From the references cited above we obtain the following statement.
\bprp\label{tExchange}
A regular tree language fulfills the exchange property if and only if it is recognized by a deterministic top-down tree automaton.
\eprp

\section{Decidability of Top-Down Determinism}\label{sDecide}
	
    It is well-known that it is decidable for a regular tree language whether or not it is top-down deterministic. 
    Viragh proved this via the construction of what he calls the {\em powerset automaton} for the path-closure of a regular language; the language is deterministic top-down, if it is equal to the language of the powerset automaton~\cite{bViragh81}.
Gécseg and Steinby used a very similar method in their textbook~\cite{bTreeAutomataBook}.
	
Another approach can be taken via an application of  Nivat and Podelksi's homogeneous closure~\cite{bNivatPodDET}. A tree language is \emph{homogeneous} if, for every $t \in L$ and every node $u \in N(t)$,
	if $t[u \leftarrow f(t_1 , t_2 )] \in L$, $t[u \leftarrow f(s_1 , t_2 )] \in L$ and also $t[u \leftarrow f(t_1,s_2 )] \in L$, then
	$t[u \leftarrow f(s_1,s_2 )] \in L$.
	The smallest homogeneous set containing a tree language is its \emph{homogeneous closure}.
	One could construct the automaton for the language's homogeneous closure. 
	The original language is deterministic top-down, if it is equal to its homogeneous closure.
	
	In both approaches the automaton of the respective closure has as state set the powerset of the original state set.
	Thus already computing this automaton takes an exponential amount of time and even space.
	The second step is in both cases the decision of the equivalence of two non-deterministic automata, which is EXPTIME-complete in the size of these automata (Corollary 1.7.9 in~\cite{bTata2007}). 
	In neither case the exact running time has been investigated.
	Also the approach of Cristau et al. \cite{bUnrankedFCT05} for unranked trees follows similar lines and does not have a better runtime.
	
	We present a new method for deciding whether a regular tree language is top-down deterministic
which runs in polynomial time.

\medskip

    In corresponding \BA{}s and \TA{}s, non-determinism in one direction corresponds to different transitions converging to the same right-hand side in the other direction. 
    We now formalize this phenomenon.  
	
	\bdf\label{tDefConflux}
	Let $A$ be a deterministic, minimal bottom-up tree automaton.
	A pair of distinct transitions $f (q_{1,1},\dots, q_{1,k} ) \ra q$ and $f(q_{2,1},\dots, q_{2,k} ) \ra q$ 
	is called a {\em conflux}.
	A maximal set of transitions, which pairwise form confluxes
(on the same input symbol $f$ and with same right-hand side $q$), is called a \emph{conflux group}.
	\edf
	
	The subtree exchange property from Definition~\ref{bExchange} essentially states that trees that appear in the same positions can be interchanged.
	For states in a \TA{} an analogous property would say that these must be exchangeable on the right-hand sides of rules;
	but this is not the case, because despite its determinism the runs for distinct occurrences of the same subtree can be distinct.
	However, when we look at the minimal deterministic bottom-up automaton for a deterministic top-down tree language, then we can establish a kind of exchange property for its states.
	
	\blem \label{tDetNoVio}
	Let $L$ be a deterministic top-down tree language and let $M$ be the minimal deterministic bottom-up automaton recognizing it. If $M$ has a conflux  of the transitions $f (q_{1,1},\dots, q_{1,k} ) \ra q$ and $f (q_{2,1},\dots, q_{2,k} ) \ra q$, then all the transitions from the set $$\{ f (q_{i_1,1},\dots, q_{i_k,k} ) \ra q\mid i_1,\dots,i_k\in \{1,2\}\}$$ are also present in $M$.
	\elem
	\bp
	Let $t_{i,1},\dots t_{i,k}$ be trees such that $t_{i,j} \in L(M_{q_{i,j}})$ for all $j\in\{1,\dots ,k\}$ and $i\in\ \{1,2\}$.
	Such trees exist by Proposition~\ref{MDFA}~\ref{MDFA1}.
	It follows from Proposition~\ref{MDFA}~\ref{MDFA2} that there exists a context $C\in \mathcal{C}(\Sigma )$ such that $C[f(t_{i,1},\dots t_{i,k})] \in L$ for  $i\in\ \{1,2\}$.
	Because the transitions $f (q_{1,1},\dots, q_{1,k} ) \ra q$ and $f (q_{2,1},\dots, q_{2,k} ) \ra q$ are distinct there exists a $j\in \{1,\dots,k\}$ such that $q_{1,j} \not= q_{2,j}$. 
	By Proposition~\ref{MDFA}~\ref{MDFA3} this implies that $t_{1,j} \not = t_{2,j}$.
	
	The tree $t = C[f(t_{1,1},\dots, t_{1,j-1}, t_{2,j} ,t_{1,j+1}, \dots , t_{1,k})]$ must be in $L$ by Proposition~\ref{tExchange}, because $L$ is deterministic top-down.
	Thus $M$ must apply a transition of the form $f (q_{1,1},\dots, q_{1,j-1 }, q_{2,j}, q_{1,j+1 },\dots, q_{1,n} ) \ra p$ for some state $p$ distinct from $q$ at the node $v$ where $f$ occurs.
	
	The two corresponding subtrees $\hat{t}_1 = t/v$ and $\hat{t}_2= \hat{t}_1[vj\leftarrow t_{2,j}]$ rooted in $v$ are not syntactically equivalent, because $M$'s states correspond to the equivalence classes of the syntactic congruence. 
	Thus there is some context $C$ such that $C[\hat{t}_1] \in L$ but $C[\hat{t}_2] \not \in L$. 
	If there is no such context, then there is one such that $C[\hat{t}_1]  \not\in L$ but $C[\hat{t}_2] \in L$, because otherwise the two trees would be syntactically equivalent; without loss of generality we treat only the former case.
	
	Because $L$ is a deterministic top-down tree language, by the exchange property from Proposition~\ref{tExchange} the tree $C[\hat{t}_2]$ should be in $L$ if $C[\hat{t}_1]$ is, since one is obtained from the other by exchanging $t_{1,j}$ for $t_{2,j}$ or the other way around, while the context $C$ remains equal. 
	This shows that no context distinguishing the trees $\hat{t}_1$ and $\hat{t}_2$ can exist, and thus $p$ must actually be equal to $q$.
	Absolutely symmetrically we can show that also  $f (q_{2,1},\dots,q_{1,j},\dots, q_{2,k} ) \ra q$ must be present in $M$. The same argument applies to each one of the $k$ positions in the conflux, which proves the statement.  
	\ep
 
 	Lemma~\ref{tDetNoVio} provides us with a necessary condition for a language to be deterministic top-down. 
 	We introduce the notion of violation for the case where the conditions of the lemma are not met.
 		
	\bdf\label{tDefViolation}
	Let $M$ be a minimal deterministic bottom-up tree automaton.  
	If there is a pair of transitions $f (q_{1,1},\dots, q_{1,k} ) \ra q$ and $f (q_{2,1},\dots, q_{2,k} ) \ra q$ in $M$ which constitute a conflux, but not all the transitions from the set $$\{ f (q_{i_1,1},\dots, q_{i_k,k} ) \ra q\mid i_1,\dots,i_k\in \{1,2\}\}$$ are also present in $M$, then we say that this conflux constitutes a {\em violation}.
	
	The transitions that form part of violations, which read the same symbol and result in the same state on the right-hand side form the corresponding {\em violating group}.
	For such a transition  $f (q_{1,1},\dots, q_{1,k} ) \ra q$ its violating group is $\{x(p_1,\dots,p_k) \ra p \in \delta \mid x=f \textrm{ and } p=q )\}$.
	\edf
	As the symbol, which is read, and the resulting state uniquely identify each violating group, each transition of a violation belongs to exactly one group.
	
	Now in the terminology of Definition~\ref{tDefViolation} the statement of Proposition~\ref{tDetNoVio} says that the \MDBA{} for a deterministic top-down tree language cannot contain any violation.
	Now we show that the absence of violations in the minimal automaton necessarily means that the language is  top-down deterministic.

\blem\label{tNoVioDet}
If the minimal deterministic bottom-up automaton $M$ for a language $L$ contains no violation, then $L$ is top-down deterministic.
\elem
\bp
If $M$ does not contain any conflux, then its corresponding \TA{} is deterministic and the statement holds.
Otherwise we construct an equivalent automaton without confluxes.
The first step in this construction is the elimination of one arbitrary conflux group. 

So let the set $\{ f (q_{i,1},\dots, q_{i,k} ) \ra q \mid i\in \{1,\dots, \ell\}\}$ be the conflux group, which we choose to eliminate, where $\ell$ is the number of transitions in this group.
We construct a new automaton without this conflux group that recognizes the same language.
\begin{itemize}
	\item Its set of states is $Q\cup \{p_j \mid j\in \{1,\dots,k\}\}$ with one new state for each position on the left-hand sides of the transitions of the conflux.
	\item We remove all the transitions of the conflux group. 
	\item Instead we add the single {\em substitute transition} $f (p_1,\dots, p_k ) \ra q$.
	\item Then for every transition $\lambda \rightarrow q_{i,j}$ that has one of the states $q_{i,j}$ on its right hand side we add the transition $\lambda \rightarrow p_j$ that has $p_j$ instead, while the left-hand side is identical; we call these copies {\em adapter transitions}.
\end{itemize}
Let $M' = (Q\cup \{p_j \mid j\in \{1,\dots,k\}\}, \Sigma, Q_f, \delta')$ be the resulting automaton, where $\delta'$ is obtained from $\delta$ by removing the conflux transitions and adding the substitution and adapter transitions as described.
In what follows we will call the components of $M$ the {\emph{original}} ones.

The idea behind this construction is the following: 
$M'$ essentially does the same runs as $M$. 
Only when $M$ applies a transition of the eliminated conflux group $M'$ applies the substitute transition instead.
In order to be able to do this, $M'$ must guess in the previous steps that instead of the original transitions applied by $M$ it should use the corresponding adapter transitions.  

\begin{clm}\label{clEq}The tree automaton $M'$ recognizes the same language as $M$. 
\end{clm}

The accepting runs of $M$ and $M'$ are in one-to-one correspondence.
This is proved in detail in the appendix.
 
\medskip

We have seen how to eliminate one conflux group.
Unfortunately, this elimination does not necessarily reduce the number of conflux groups. 
If the state $q'$, which is on the right-hand side of all rules of a different conflux group, appears as one of the $q_{i,j}$ in the eliminated conflux group, then a copy of the entire group with $q'$ is made in the adapter transitions. 
Note that also the newly introduced $f (p_1,\dots, p_k ) \ra q$ could be a transition with one of the $q_{i,j}$ on its right hand side if this  $q_{i,j}$ is equal to $q$, see also Example~\ref{eQRecurs}.
In this case, however, the conflux group is not copied, because its transitions are removed before the adapter transitions are introduced. 

So the number of conflux groups can stay the same and even increase. 
Nonetheless we start by removing all the original conflux groups in the way described.

\begin{clm}\label{NoCopyCopy}
	After all the original conflux groups are removed, further removals always decrease the number of conflux groups. 
\end{clm}

New conflux groups can only be added in the step where the adapter transitions are introduced, because the substitute transition obviously creates no new conflux group.
So all non-original conflux groups consist of adapter transitions and thus do not have states from the original $Q$ on their right-hand sides.
But these new states never occur on the left-hand side of any transition except their corresponding substitute transitions, which cannot form part of any conflux.
Consequently they are never copied for new adapter transitions.
Therefore only original conflux groups can be copied and Claim~\ref{NoCopyCopy} holds.

Summarizing, we do one elimination step for each original conflux group.
After this a number of copies of original conflux groups can have appeared.
During their elimination their number decreases by one in every step.
Therefore this process terminates and we obtain a bottom-up automaton, which does not have any confluxes and is equivalent to the orignal \MDBA.
It is not deterministic anymore, but now its corresponding \TA{} is, because only confluxes result in nondeterministic choices in the reversal.
\ep	

We illustrate the construction in the proof of Lemma~\ref{tNoVioDet} with two examples.

\bex
Consider the language $$L = \{ f(a, f (a,b) ) ,  f(a, f (b,a) ) ,  f(a, f (a,a) ) ,  f(a, f (b,b) )\}.$$ 
It is deterministic top-down, but its \MDBA{} contains a conflux, which is not a violation.
Its transitions are $q_0(a) \ra q_a$, $q_0(b) \ra q_b$, $f(q_a, q_b) \ra q$,  $f( q_b, q_a) \ra q$, $f( q_a, q_a) \ra q$, $f( q_b, q_b) \ra q$, and $f( q_a, q) \ra q_f$. 
The four transitions with $q$ on the right-hand side constitute a conflux but not a violation. 

Applying the construction we introduce the new states $p_1$ and $p_2$.
The four transitions of the conflux are replaced by the substitute transition $f( p_1, p_2) \ra q$.
Further the adapter transitions $q_0(a) \ra p_1$, $q_0(b) \ra p_1$, $q_0(a) \ra p_2$, and  $q_0(b) \ra p_2$ are added.
The resulting automaton has the same number of transitions and two additional states.
In this case $q_b$ could be deleted, because it can only be read by the transitions of the conflux; in general original states do not become obsolete as shown by $q_a$.
The recognized language is the same, but on leaves labeled $a$ there is the non-deterministic choice of going into state $q_a$, $p_1$, or $p_2$, similarly for leaves labeled by $b$.
\eex

\bex\label{eQRecurs}
An interesting case for the construction in the proof of Lemma~\ref{tNoVioDet} is the occurence of a state on both the left-hand and the right-hand side of a transition of the conflux. 
Let $f(q',q) \ra q$ be such a transition. 
When it (along with the other ones) is removed, it is replaced by $f(p_1,p_2) \ra q$.
Then also $f(p_1,p_2) \ra p_2$ is added.
If we did the latter step before adding $f(p_1,p_2) \ra q$, then this recursivity would be lost.
So the order in which transitions are added and removed is essential.
\eex

Together, Lemmas~\ref{tNoVioDet} and~\ref{tDetNoVio} provide us with a characterization of the deterministic top-down tree languages.

\bthm\label{tDTDNoVio}
A regular tree language is top-down deterministic if and only if its minimal deterministic bottom-up automaton contains no violations.  
\ethm

This provides us with a new method to decide whether a regular tree language $L$ given as a deterministic bottom-up automaton is top-down deterministic:
\begin{enumerate}
	\item\label{StepMinim} Compute the minimal deterministic bottom-up tree automaton $M$ for $L$.
	\item\label{StepConflux} Find all confluxes in $M$'s set of transitions.
	\item\label{StepDecide} For each conflux check whether it constitutes a violation.
\end{enumerate}	

Step~\ref{StepMinim} can be computed in quadratic time. 
Carrasco et al.~\cite{bMinimizeQuadratic} showed in detail how to minimize a deterministic bottom-up automaton within this time bound. 
Minimization algorithms were already known early on, but their runtime was not analyzed in detail~\cite{bBrainerd,bTreeAutomataBook}.

Both Steps~\ref{StepConflux} and~\ref{StepDecide} are purely syntactical analyses of the set of transitions.
To optimize the runtime we can group 
the strings describing transitions into classes $T(q_f) = \{(q_1,\dots,q_k) \mid f(q_1,\dots,q_k) \ra q \in \delta\}$ for all states $q$ and all node labels $f$ in linear time in the style of bucket sort.
The different transitions of a possible conflux group are all in the same class which, on the other hand, is not longer than the total description of the automaton.
Thus linearly many transitions need to be compared in order to determine whether there is a conflux and whether it constitutes a violation.
Also this takes an amount of time at most quadratic in the size of the input.

\bthm
For a regular tree language given as a \DBA{} it is decidable in quadratic time whether it is also deterministic top-down.
\ethm

\section{Finite Unions of Deterministic Top-Down Tree Languages}

In the preceding section we have provided a new characterization of the class of top-down deterministic tree languages.
One deficiency of this class is that it is not closed under basic operations such as set-theoretic union. 
Moreover, even simple finite languages such as $\{f(a,b), f(b,a)\}$ are not included in this class.
To remedy these deficiencies, we consider 
finite unions of top-down deterministic tree languages.
They contain many common examples for non-top-down deterministic tree languages, 
but still are characterized by deterministic top-down tree automata.
We denote this class by $\FUDT$.

In Section~\ref{sDecide} we have seen that violations in the minimal deterministic bottom-up automaton can be used to decide whether a language is deterministic top-down.
Among the automata with violations, some recognize languages that are still in $\FUDT$ while other ones recognize languages outside this class.
We now explore how an analysis of the occurring violations can be used to determine to which one of the classes a given language belongs.  
 
To this end we use a context-free grammar $G(M)$ to analyze where and how a given \MDBA{} $M$ uses the transitions of its violations. 
This \emph{violation grammar} 
\begin{itemize}
	\item has $M$'s state set plus a new start symbol $S$ as its set of non-terminals.
	\item The terminals are $[$, $]$ and one distinct \emph{violation symbol} for each of the violating groups of $M$.
	\item For every transition $f(q_1, \dots, q_k) \ra q$ from a violating group $\nu$ we add the production $ q \ra \nu[ q_1 \cdots q_k]$;
	\item for all transitions that are not from any violating group we add the production $ q \ra  q_1 \cdots q_k$.
	This implies that for initial states $q_0$ there are rules $q_0\ra \epsilon$.
	\item Finally, there is the transition $S \ra [q_f]$ for each final state $q_f$ of $M$.
\end{itemize}
For a run $\beta$ of $M$ we call its \emph{corresponding string} $\theta(\beta)$ the terminal string that is generated by $G$ by using the productions corresponding to the transitions used in $\beta$ in the corresponding order.
The \emph{violation tree} of $\beta$ is obtained from $\theta(\beta)$ as follows:
All brackets $[ \ ]$ without any other non-terminals between them are removed from $\theta(\beta)$.
A root note is introduced, and then the bracket structure is translated to a tree in the natural way.
For example, a string $[\eta_1[\ ]\eta_2[\eta_1[\ ]\ ]\ ]$ results in the tree $\epsilon(\eta_1,\eta_2(\eta_1))$.
Nodes with symbols of violations are called \emph{violation nodes}.

  \blem\label{tMaxLength}
   Let $M$ be an \MDBA{} and let $n$ be the number of transitions that form part of violations and are applied in the run $\beta$ of $M$ on a tree $t$. 
 Then the corresponding string $\theta(\beta)$ for this run has length $3n+2$.  
  \elem
  \bp
  The unique production for the start state adds two terminals, namely $[$ and $]$.
  The only other productions that generate terminals are the ones corresponding to transitions that form part of violations.
  Each one adds three terminals one of which is a violation symbol.
  \ep
So the language $L(G(M))$ is finite if and only if there is some number $n$ such that every accepting run of $M$ uses at most $n$ times transitions that form part of some violation.

\blem\label{tMapInf}
Let $M$ be an \MDBA{}.
If $M$'s violation grammar $G(M)$ generates an infinite language, then $L$ is not in $\FUDT$.
\elem
Essentially, every violation symbol represents a choice that cannot be made in a top-down deterministic way.
All of these choices are pairwise independent in the sense that for each one a new \DTA{} is necessary.
So if there is no bound on their number, no finite union can be found. 
The technical details can be found in the appendix.

If the violation grammar's language is not infinite as in Lemma~\ref{tMapInf}, then we can construct a family of \DTA{}s that demonstrate that the given language is in $\FUDT$.
  
 \blem \label{tMapNotInf}
 Let $M$ be an \MDBA{}.
 If $M$'s violation grammar produces a finite language, then $L$ is a finite union of deterministic top-down tree languages.
 \elem
 \bp
 We first treat the case where $L(G(M))$ is a singleton set.
 If the violation tree contains at most one violation node per group, then we decompose the \MDBA{} $M$ as follows: 
 for every possible combination of transitions from the violating groups that contains exactly one transition from each group we make one automaton that contains exclusively these transitions from the respective violating groups.  
 In addition it contains all the other transitions that do not belong to any violating group.
 The total number of automata is $\prod\{|\eta| \mid \eta \textrm{ is a violating group in } M\}.$
 
 These automata do not contain violations any more, because all the existing ones have been removed and no new transitions have been added. 
 Thus their corresponding \TA s are deterministic top-down automata or can be transformed as in the proof of Lemma~\ref{tNoVioDet} by eliminating all confluxes.
 Finally, let $K$ be the union of the languages of all the new automata.
 $K\subseteq L$, because every run in one of the new automata can be done by exactly the same transitions in $M$; on the other hand, also for every run of $M$ there is one new automaton that contains all the transitions that are used, and thus $L\subseteq K$ and consequently $L=K$. 
 So we have decomposed $L$ into a union of deterministic top-down tree languages.
 
 From the proof of Lemma~\ref{tMapInf} we can see that for every pair $(L_1,L_2)$ of these languages there is a pair of trees that show that $L_1\cup L_2$ can never be part of a deterministic top-down subset of $L$.
 Thus there cannot be any decomposition with fewer components.
 
We only sketch how to generalize this construction to several occurrences of the same violating group in the string and then to $L(G(M))$ consisting of several strings.  
 If some violating group $\nu$ appears several times in the string $s$, at each occurrence of $\nu$ a different transition from $\nu$ could be used in a run of $M$.
 So instead of choosing one fixed transition from the group, we independently choose one for each occurrence and with it its position in the tree; we index the transition with the position of the occurrence in the violation tree.
 When the new automaton applies one of these transitions, it remembers its position and verifies it, while moving up in the input tree.
 Similarly, occurrences of $\nu$ in distinct strings can be distinguished.
 Appendix \ref{a14} explains this in more detail.
 \ep

 The number of automata introduced in the proof of Lemma~\ref{tMapNotInf} is exponential in the number of nodes in the violation grammar's output language.
 This might seem bad at first sight;
 however, from the proof of Lemma~\ref{tMapInf} we can see that for a single tree in the output language this number cannot be improved.

 \bthm\label{tDecideFinUn}
 For a regular tree language given as a \DBA{} $M$ it is decidable in quadratic time whether or not it belongs to the class $\FUDT$. 
 \ethm
 \bp
 We proceed as follows:
 \begin{enumerate}
 	\item\label{D1} Construct the minimal deterministic bottom-up automaton $M'$ for $L$.
 	\item\label{D2} Detect all violations in $M'$. 
 	\item\label{D3} Construct the violation grammar for $M'$.
 	\item\label{D4} Decide whether the grammar's language is finite.
 \end{enumerate}
 Steps~\ref{D1} and~\ref{D2} are just as in the procedure following Theorem~\ref{tDTDNoVio}. 
 The construction of the violation grammar has been described above.
 Now the question of Step~\ref{D4} is equivalent to our decision problem by Lemmas~\ref{tMapInf} and~\ref{tMapNotInf}.
 For this decision we first eliminate all deleting rules from the grammar, which can be done in linear time~\cite{bHarYehu}. 
 With this reduced grammar the finiteness of the language can be decided essentially by detecting cycles in the transition graph.
 This can be done in time linear in the number of edges and nodes of the graph by detecting the \emph{strongly connected components (SCC)}~\cite{bTarjan72}.
 If in any SCC a rule is used that produces more than one non-terminal, then the grammar's language is infinite, otherwise it is not. 
 Also this check and therefore the entire Step~\ref{D4} can be done in linear time.
 \ep

\section{Conclusions}
	
The concept of violations in minimal deterministic bottom-up tree automata 
allows to decide whether the given language is top-down deterministic, or
a finite union of top-down determinsitic tree languages.
In the affirmative cases, corresponding represenations using those formalism
can be constructed, but in the case of finite unions may be exponential in size.

	\bibliographystyle{splncs04}
	\bibliography{PTL}

	\newpage
	\appendix 
	
	\section{Proof of Claim~\ref{clEq} in the Proof  of Lemma~\ref{tNoVioDet}}

	$L(M)\subseteq L(M')$, because every accepting run in $M$ for a tree $t$ has a corresponding accepting run for $t$ in $M'$. 
	As described above, any application of a rule from the removed conflux group can be simulated by the adapter and substitute transitions.
	All the other transitions from $\delta$ are also present in $\delta'$.
	
	Similarly $L(M')\subseteq L(M)$ holds because of a one-to-one correspondence between accepting runs.
	We first point out three consequences of the way in which we construct $M'$:
	\begin{enumerate}
		\item\label{A1} The states from $\{p_j \mid j\in \{1,\dots,k\}\}$ are read exclusively by the substitute transition or an adapter copy thereof.
		\item\label{A2} $\bigcup_{i\in \{1,\dots,\ell\}} L(M'_{q_{i,j}}) = L(M'_{p_j})$, i.e., whenever $M'$ reads a tree using an adapter transition to $p_j$ last, it can read the same tree using a original transition to one of the corresponding $q_{i,j}$ in the last step instead.
		\item \label{A3} If $L(M'_{q_{{i_j},j}}) \subseteq L(M'_{p_j})$ for all $j \in \{1,\dots, k\}$ and $i_j \in \{1,\dots, \ell\}$, then the transition $f (q_{i_1,1},\dots, q_{i_k,k} ) \ra q$ is in $\delta$.
	\end{enumerate}
	Let $\beta'$ be an accepting run for a tree $t$ in $M'$.
	If $\beta'(u) = p_i$ for some node in $t$, by~\ref{A2} we can find an original state $q_{i,j}$ such that $t/u\in L(M'_{q_{i,j}})$.
	All the siblings of $u$ must be mapped to states from $\{p_j \mid j\in \{1,\dots,k\}\}$, because otherwise from~\ref{A1} we can see that $\beta'$ could not continue;
	more precisely, the $m$-th sibling must be mapped to $p_m$ for $m\in \{1,\dots,k\}$, because from~\ref{A1} we know that $f (p_1,\dots, p_k )$ is the right-hand side of all rules that read the new states.
	Also for these siblings of $u$, we can find original states analogous to $q_{i,j}$, for which by~\ref{A3} there exists a fitting transition $f (q_{i_1,1},\dots, q_{i_k,k} ) \ra q$ in $\delta$.
	In this way $\beta'$ can be changed to become an accepting run of $M$.
	\qed

	\section{Proof of Lemma~\ref{tMapInf}}
	
	Let us suppose that there is a number $\ell$ such that $L$ is the union of $\ell$ deterministic top-down tree languages while $L$'s violation grammar generates an infinite language.
	Because the language is infinite, there is no bound on the length of its strings.
	From Lemma~\ref{tMaxLength} we can see that in this case there is no bound on the number of violation symbols in strings of this language either.
	Let $s$ be the violation tree of such a string with more than $\ell$ violation symbols.
	
	Now let $u'$ and $v'$ be two distinct nodes other than the root in $s$.
	The violating group for $u'$ contains at least two distinct transitions.
	Let $t_u$ and $t_u'$ be two trees that are read by $M_L$ using one of these transitions (each of the two a distinct one) in the last step, i.~e., arriving at the root.
	We take an arbitrary tree $t\in L$ whose run corresponds to $s$ by the violation grammar;
	$u$ and $v$ are the nodes in $t$, which correspond to $u'$ and $v'$ in $s$, respectively.  
	Then $t[u\leftarrow t_u]$ and $t[u\leftarrow t_u']$ are both in $L$, because the accepting run for $t$ in $M_L$ can be translated to accepting runs for these two trees; one of the two trees might actually be equal to $t$.
	
	However, we know that the two trees can never be in the same deterministic top-down subset of $L$:
	$u$ corresponds to a violating group and in this group at least one transition of the ones that would turn it into a conflux that is not a violation is missing.  
	Therefore there is a tree $t_u''$ with the following properties:  
	\begin{itemize}
		\item $t_u''$ is obtained in the following way: let $f(q_1,\dots, q_k) \ra q$ be the missing transition.  
		In $t_u$, let $t_1$ to $t_k$ be the root's children in numerical order.
		For each $j\in\{1,\dots,k\}$, if $t_j\not\in L(M_{q_j})$ then we substitute $t_j$ by a tree from $L(M_{q_j})$.
		\item  $t[u\leftarrow t_u'']$ is not in $L$.
	\end{itemize}
	So with the missing transition $M$ could read $t_u''$, but without it this is not possible.
	We have already seen in Section~\ref{sDecide} that such a tree must exist, because a violating group always leads to a violation of the exchange property from Definition~\ref{bExchange}.
	
	For the node $v$ we can find analogous trees $t_v$, $t_v'$ and $t_v''$ such that   $t[v\leftarrow t_v]$ and $t[v\leftarrow t_v']$ are both in $L$ but can never be in the same deterministic top-down subset of $L$, because $t[v\leftarrow t_v'']$ would have to be in this set to fulfill the exchange property.
	From this we can conclude that the four trees  $t[u\leftarrow t_u, v\leftarrow t_v]$, $t[u\leftarrow t_u', v\leftarrow t_v]$, $t[u\leftarrow t_u, v\leftarrow t_v']$, and $t[u\leftarrow t_u', v\leftarrow t_v']$ can pairwise never be in the same deterministic top-down subset of $L$.
	
	If there is a descendancy relation between $u$ and $v$, then it might not be possible to conduct the two  substitutions simultaneously. 
	Let, without loss of generality, $v$ be a descendant of $u$.  
	Then it is important to chose one of the two trees $t_u$ and $t_u'$, say $t_u$ in such a way that $t[u\leftarrow t_u]$ is the original $t$ and thus contains also $v$. 
	In this case we obtain the three trees $t[u\leftarrow t_u, v\leftarrow t_v]$, $t[u\leftarrow t_u, v\leftarrow t_v']$, and $t[u\leftarrow t_u']$, which can pairwise never be in the same deterministic top-down subset of $L$.
	
	From the two violation nodes $u'$ and $v'$ we have obtained at least three trees that can pairwise never be in the same deterministic top-down subset of $L$. 
	Because there are more than $\ell$ nodes labeled by violating groups in $s$, we can find more than $\ell$ ($2^\ell$ if there is no descendancy relation between the $u$ and $v$, or if $v$ can be produced by different transitions in the $u$) trees from $L$, which can pairwise not be in the same deterministic top-down subset of $L$.  
	But this contradicts our initial assumption that there is a number $\ell$ such that $L$ is the union of $\ell$ deterministic top-down tree languages, and we can conclude that $L$ is not in $\FUDT$.
	\qed

	\section{Details on the Proof of Lemma \ref{tMapNotInf}}\label{a14}
	
	We have seen how to construct a family of \DTA{}s that recognize a set of languages such that their union is $L$ in the case where $L(G(M))$ is a singleton set, and the violation tree derived from the string in this set contains at most one violation node per group.
	Essentially this was done by deleting all but one transition from each violating group.
	This amounts to choosing, which transition from this group will be applied at the unique position, where it can be applied.
	If there are more than one positions, we cannot make this fixed choice for all of them, because in each position a different transition can be applied.
	Therefore we introduce one copy of the violating group for each position and check, whether the transitions are applied only at the respective positions. 
	
	Let $L(G(M))$ be a singleton set, where the tree in this set contains an arbitrary number of violation nodes per group.
	We first modify the \MDBA{} so that it contains one copy of the respective violating group for each violation node. 
	Then we can apply the same technique of separating the deterministic parts of the automaton as above. 
	Let $P$ be the set of positions in the violation tree and let $P(\eta)$ be the subset of all the positions labeled with $\eta$.
	We equip the \MDBA{}'s states with a buffer that is empty, when a leaf is read.
	Transitions that do not form part of a violating group produce their right-hand side with empty buffer, if all the buffers on the left-hand side are empty. 
	If there are positions in the buffers on the left-hand side, we check whether these are
	\begin{itemize}
		\item all of the same depth and
		\item in order.
	\end{itemize}
	If this is the case, then the state of the right-hand side is produced with the ordered list of all these positions in its buffer.
	
	For transitions that form part of a violating group $\eta$, if all the buffers on the left-hand side are empty we guess a leaf position from $P(\eta)$ and put it in the buffer.
	If there are positions in the buffers on the left-hand side, again we check whether they are of the same depth, say $d$ and in order.
	Further, we check whether the corresponding parent node $u$ of depth $d-1$ in the violation tree is labeled by $\eta$ and whether the positions in the buffers represent the complete list of children of $u$.
	Only if all of this is the case, we produce the state on the right-hand side with only the position of $u$ in its buffer.
	
	The only final state is the original final state with the complete list of positions on level one of the violation tree in its buffer.
	This new tree automaton $A$ recognizes the same tree language as the original  \MDBA{}.
	Every accepting run of $A$ can be converted to an accepting run of the  \MDBA{} by just deleting the buffer from the states;
	on the other hand, for every accepting run of the  \MDBA{} there is a series of guesses that fill the buffers in such a way that the final state is reached.
	
	\bex\label{e8}
	We illustrate the construction with an example: we consider the tree language which consists of the eight trees 
	\begin{align*}
		f(f(a,b),f(f(a,b),f(a,b)))\\
		f(f(a,b),f(f(a,b),f(b,a)))\\
		f(f(a,b),f(f(b,a),f(a,b)))\\
		f(f(a,b),f(f(b,a),f(b,a)))\\
		f(f(b,a),f(f(a,b),f(a,b)))\\
		f(f(b,a),f(f(a,b),f(b,a)))\\
		f(f(b,a),f(f(b,a),f(a,b)))&\\
		f(f(b,a),f(f(b,a),f(b,a)))&.
	\end{align*}   
	So in three positions there is a choice between $(a,b)$ or $(b,a)$ as leaf children.
	These result in applications of transitions of the same violating group in three different positions. 
	That is why the violation tree $\epsilon(\eta, \eta, \eta)$ has three occurrences of $\eta$, if this is the corresponding symbol.
	
\begin{figure}[h]
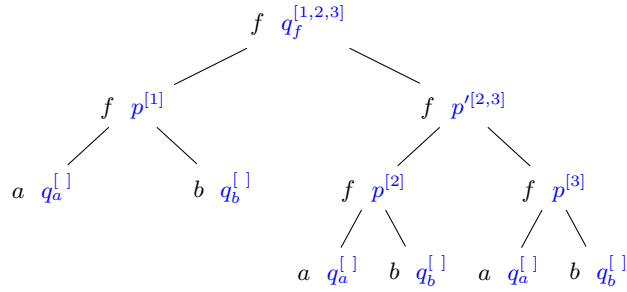
 \centering
		\raisebox{-.5\height}{\tikz[
			level 1/.style={sibling distance=44mm},
			level 2/.style={sibling distance=24mm},
			level 3/.style={sibling distance=12mm}, level distance = 11mm]{
				\node {$f \ \ \textcolor{blue}{q_f^{[1,2,3]}}$} 
				child { node {$f \ \ \textcolor{blue}{p^{[1]}}$}
					child {node {$a \ \ \textcolor{blue}{q_a^{[\ ]}}$}}
					child {node {$b \ \ \textcolor{blue}{q_b^{[\ ]}}$}}
				}
				child { node {$f \ \ \textcolor{blue}{p'^{[2,3]}}$}
					child { node {$f \ \ \textcolor{blue}{p^{[2]}}$} 
						child {node {$a \ \ \textcolor{blue}{q_a^{[\ ]}}$}}
						child {node {$b \ \ \textcolor{blue}{q_b^{[\ ]}}$}}
					}
					child { node {$f \ \ \textcolor{blue}{p^{[3]}}$}
						child {node {$a \ \ \textcolor{blue}{q_a^{[\ ]}}$}}
						child {node {$b \ \ \textcolor{blue}{q_b^{[\ ]}}$}}
					}
				}
				;
			}
		}
	\caption{An accepting run.
		At the side of each node we see in \textcolor{blue}{blue} the state reached after reading this node.
		 The states  have the buffer in the brackets in the exponent.}
	 \label{eBuffer}
	\end{figure}

	In detail, the \MDBA{} has the transitions:
	$$a\ra q_a, b\ra q_b, f(q_a,q_b)\ra p, f(q_b,q_a)\ra p, f(p,p)\ra p', f(p,p')\ra q_f$$
	where $q_f$ is the only final state.
	$f(q_a,q_b)\ra p$ and $f(q_b,q_a)\ra p$ constitute the violating group $\eta$, and $P(\eta) = \{1,2,3\}$, where all of these positions are leaves.
	An accepting run of the new tree automaton is depicted in Figure \ref{eBuffer}.
	Note how positions of the same level in the violation tree do not necessarily correspond to positions of the same level in the tree that is recognized.  
\eex

	The new automaton has one copy of the respective violating group for each violation node, namely the one with the node's position in the buffer of the state on the right-hand side of the transitions.
	Now we can obtain a family of top-down deterministic tree languages via the automata that are obtained by chosing only one transition of each group as above.
	
	If the violation grammar's language contains several strings, then we construct separate sets of automata for them.
	Here the decomposition might not be optimal in the number of automata any more. 	
	
\begin{figure}[h]
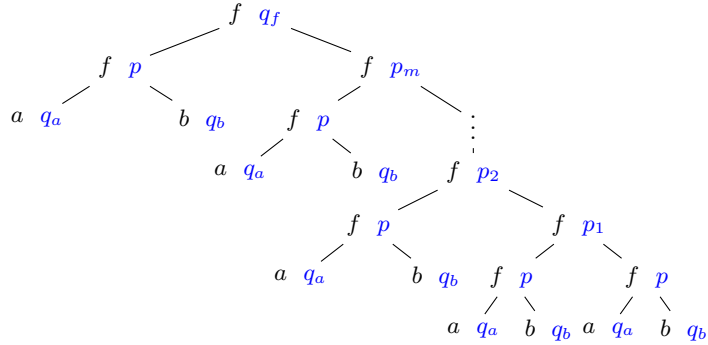
 \centering
		\raisebox{-.5\height}{\tikz[
			level 1/.style={sibling distance=36mm},
			level 2/.style={sibling distance=22mm},
			level 3/.style={sibling distance=18mm},
			level 4/.style={sibling distance=28mm},
			level 5/.style={sibling distance=18mm},
			level 6/.style={sibling distance=10mm},
			level 7/.style={sibling distance=10mm}, level distance = 7mm]{
				\node {$f \ \ \textcolor{blue}{q_f}$} 
				child { node {$f \ \ \textcolor{blue}{p}$}
					child {node {$a \ \ \textcolor{blue}{q_a}$}}
					child {node {$b \ \ \textcolor{blue}{q_b}$}}
				}
				child { node {$f \ \ \textcolor{blue}{p_m}$}
					child { node {$f \ \ \textcolor{blue}{p}$} 
						child {node {$a \ \ \textcolor{blue}{q_a}$}}
						child {node {$b \ \ \textcolor{blue}{q_b}$}}
					}
					child { node {$\vdots$} 
						child { node {$f \ \ \textcolor{blue}{p_2}$}
							child { node {$f \ \ \textcolor{blue}{p}$}
								child {node {$a \ \ \textcolor{blue}{q_a}$}}
								child {node {$b \ \ \textcolor{blue}{q_b}$}}
							}
							child { node {$f \ \ \textcolor{blue}{p_1}$}
								child { node {$f \ \ \textcolor{blue}{p}$}
									child {node {$a \ \ \textcolor{blue}{q_a}$}}
									child {node {$b \ \ \textcolor{blue}{q_b}$}}
								}
								child { node {$f \ \ \textcolor{blue}{p}$}
									child {node {$a \ \ \textcolor{blue}{q_a}$}}
									child {node {$b \ \ \textcolor{blue}{q_b}$}}
								}
							}
						}
					}
				}
				;
			}
		}
		\caption{A family of trees which requires exponentially many elements in its decomposition into deterministic top-down tree languages.}
		\label{eExpo}
	\end{figure}

	\bex
	We generalize Example \ref{e8} to show how big the number of \DTA{} can become with respect to the number of states of the \MDBA.
	Again the tree language consists of a fixed backbone along which there are choices between the subtrees $f(a,b)$ and $f(b,a)$ as in Example \ref{e8};
	only the depth can be greater and is parameterized by the integer $m$.
	Figure \ref{eExpo} depicts an example tree and again in blue the states of the accepting.
	The \MDBA{}'s state set is $\{ q_a,q_b, p, q_f, p_1, p_2, \dots, p_m \}$ and has $m+4$ elements.
	The transitions should be evident from Figure \ref{eExpo} together with Example \ref{e8}.

	The violation is the same one as in Example \ref{e8}.
	So for every use of the state $p$ two copies of this violation are created.
	There are $m+2$ nodes with the state $p$ and consequently the number of tree automata that are created is $2^{m+2}$.
	Each of these recognizes only one tree.
	In the same way as in the proof of Lemma \ref{tMapInf} we can see that these can pairwise never be in the same deterministic top-down subset of the language. 
	Thus there cannot be any decomposition with fewer elements.
	\eex

	\section{Details on the Proof of Theorem \ref{tDecideFinUn}}
	
	We look in more detail at Step~\ref{D4} of deciding whether the language of the violation grammar is finite.
	First off, we point out that the grammar is reduced in the sense that there are no unreachable and only productive non-terminals.
	This is due to the construction and the fact that the underlying \MDBA{} has no useless or unreachable states;
	specifically we excluded the sink state in its definition, which corresponds to the class of trees, which can never be subtrees of any tree of the language.
	
	After eliminating the deleting rules, i.e. rules with the empty string on the right-hand side, we construct a graph for the detection of cycles.
	This graph has the grammar's non-terminals as nodes.
	The directed edges are derived from the productions.
	For all chain productions $A\rightarrow B$ we add the corresponding edge.
	For all productions whose right-hand sides have a length greater than one, we add edges to all the non-terminals that occur on the right-hand side. 
	These edges are marked so that the algorithm for detecting cycles can recognize them. 
	
	If any of these marked edges is used in a strongly connected component, this means that the corresponding rule can be applied an arbitrary number of times in a derivation of the grammar.
	Because every application lengthens the string and there are no deleting rules, there is no bound on the length of strings that are generated by the grammar.
	Consequently the language must be infinite.
\end{document}